\documentstyle[11pt,aaspp4]{article}
\journalid{}{}
\articleid{}{}
\slugcomment{}
\begin{document}

\def\plotthree#1#2#3{\centering \leavevmode
\epsfxsize=0.30\columnwidth \epsfbox{#1} \hfil
\epsfxsize=0.30\columnwidth \epsfbox{#2} \hfil
\epsfxsize=0.30\columnwidth \epsfbox{#3}}

\title{A Kinematic Link between Boxy Bulges, 
Stellar Bars, and Nuclear Activity in NGC~3079 \& NGC~4388}

\author{S. Veilleux\altaffilmark{1,2,3}, J.
Bland-Hawthorn\altaffilmark{3,4}, and G. Cecil\altaffilmark{5}}

\altaffiltext{1}{Department of Astronomy, University of Maryland, College Park,
MD 20742; E-mail: veilleux@astro.umd.edu}

\altaffiltext{2}{Cottrell Scholar of the Research Corporation}

\altaffiltext{3}{Visiting Astronomers, Canada-France-Hawaii Telescope,
operated by the National Research Council of Canada, the Centre de la 
Recherche Scientifique de France, and the University of Hawaii }

\altaffiltext{4}{Anglo-Australian Observatory, P.O. Box 296, Epping, 
NSW 2121, Australia; E-mail: jbh@aaoepp2.aao.gov.au}

\altaffiltext{5}{SOAR/NOAO, 950 N. Cherry Ave., Tucson, AZ 85726-6732;
email: gcecil@noao.edu}

\begin{abstract}
We present direct kinematic evidence for bar streaming motions in two
active galaxies with boxy stellar bulges. The Hawaii Imaging
Fabry-Perot Interferometer was used on the Canada-France-Hawaii 3.6-m
telescope and the University of Hawaii 2.2-m telescope to derive the
two-dimensional velocity field of the line-emitting gas in the disks
of the Sc galaxy NGC~3079 and the Sb galaxy NGC~4388. In contrast to
previous work based on long-slit data, the detection of the bar
potential from the Fabry-Perot data does not rely on the existence of
inner Lindblad resonances or strong bar-induced shocks. Simple
kinematic models which approximate the intrinsic gas orbits as
nonintersecting, inclined elliptical annuli that conserve angular
momentum characterize the observed velocity fields.  In NGC~3079, bar
streaming motions with moderately eccentric orbits ($e$ = $b/a$ $\sim$
0.7) aligned along PA = 130$\arcdeg$ intrinsic to the disk (PA =
97$\arcdeg$ on the sky) are detected out to $R_b$ = 3.6 kpc. The
orbits become increasingly circular beyond that radius ($e$ = 1 at
$R_d$ $\approx$ 6 kpc).
The best model for NGC~4388 includes highly eccentric orbits ($e$
$\sim$ 0.3) for $R_b \la 1.5$ kpc which are aligned along PA =
135$\arcdeg$ intrinsic to the disk (PA = 100$\arcdeg$ on the sky).
The observed ``spiral arms'' are produced by having the orbits become
increasingly circular from the ends of the bar to the edge of the disk
($R_d$ $\approx$ 5 kpc), and the intrinsic bar PA shifting from
135$\arcdeg$ to 90$\arcdeg$.

Box-shaped bulges in both NGC~3079 and NGC~4388 are confirmed using
new near-infrared images to reduce dust obscuration.  Morphological
analysis of starlight in these galaxies is combined with the gas
kinematics derived from the Fabry-Perot spectra to test evolutionary
models of stellar bars that involve transitory boxy bulges, and to
quantify the importance of such bars in fueling active nuclei.  Our
data support the evolutionary bar models, but fail to prove
convincingly that the stellar bars in NGC~3079 and NGC~4388 directly
trigger or sustain the nuclear activity.
\end{abstract}


\section{Introduction}

The origin of galaxy nuclear activity\footnote{In this paper, `nuclear
activity' refers to either starburst or black-hole driven activity in
the nuclei of galaxies. Similarly, we refer to `active galaxies' as
galaxies powered by star formation (starburst galaxies) or through
accretion onto a massive black hole (active galactic nuclei or
quasars).} is of fundamental astrophysical importance.  Recent surveys
suggest that a nonaxisymmetric component to the gravitational
potential is necessary to start nuclear activity (e.g., Moles,
M\'arquez, \& P\'erez 1995).  Evidence has mounted that galaxy
interactions trigger activity in high-luminosity galaxies (e.g.,
Sanders \& Mirabel 1996; Bahcall et al. 1997; Stockton 1998).
However, the evidence is less convincing in lower luminosity,
interacting spiral galaxies. These are known to have higher star
formation rates on the average than isolated galaxies (e.g., Kennicutt
\& Keel 1987; Keel \& van Soest 1992), but the role of galaxy
interactions in triggering nuclear activity in Seyfert galaxies is
debated vigorously (e.g., Dahari 1984; Fuentes-Williams \& Stocke
1988; Dultzin-Hacyan 1998; De~Robertis, Yee, \& Hayhoe 1998).

In low-luminosity active galaxies stellar bars may funnel gas down to
the scale of the central engine.  Observations (e.g., Quillen et
al. 1995; Benedict, Smith, \& Kenney 1996; Regan, Vogel, \&
Teuben 1997) and numerical simulations (e.g., Athanassoula 1992;
Friedli \& Benz 1993, 1995; Piner, Stone, \& Teuben 1995) have shown
that stellar bars can induce mass inflow at rates sufficient to fuel
kpc-scale starbursts ($\ga$ 1 M$_\odot$ yr$^{-1}$). These results are
consistent with the statistical excess of barred galaxies among
starbursts (Hawarden et al. 1986; Dressel 1988; Arsenault 1989; Martin
1995; Ho 1996; Huang et al. 1996; see Pompea \& Rieke 1990 and Isobe
\& Feigelson 1992, however).  It is unclear how these significant
inflow rates can be sustained down to the nuclear scale to feed the
AGN, but simulations of ``bars within bars'' (Norman \& Silk 1983;
Shlosman, Frank, \& Begelman 1989; Wada \& Habe 1992; Friedli \&
Martinet 1993; Heller \& Shlosman 1994; Maciejewski \& Sparke 1997)
and the detections of nested bars (Shaw et al. 1995; Wozniak et
al. 1995; Friedli et al. 1996; Erwin \& Sparke 1998) and ``nuclear
mini-spirals'' (e.g., Ford et al. 1994; Regan \& Mulchaey 1999;
Martini \& Pogge 1999) are promising developments.

Despite these efforts, it is surprising to find little or no
observational evidence for Seyfert nuclei to occur preferentially in
barred systems (e.g., McLeod \& Rieke 1995; Heraudeau et al. 1996;
Mulchaey
\& Regan 1997) or for emission-line strengths of AGNs to depend on the
presence of a bar (Ho, Filippenko, \& Sargent 1997).  Perhaps nuclear
activity in unbarred galaxies was triggered by short-lived bars that
have since disappeared.
Indeed, evolutionary models of disk galaxies suggest that
stellar bars are transient features that form and dissolve over
only a few orbital periods (e.g., Hohl \& Zang 1979;
Miller \& Smith 1979; Combes \& Sanders 1981; Combes et al. 1990;
Pfenniger \& Friedli 1991; Raha et al. 1991; Merritt \& Sellwood 1994;
Norman, Sellwood, \& Hasan 1996). In these scenarios, a vertical
instability in the bar kicks stars above the disk of the galaxy to
produce boxy peanut-shaped bulges that eventually settle to become
stellar spheroids.  Recent kinematic evidence for stellar bars
in galaxies with boxy bulges has provided observational support
for this scenario (e.g., Bettoni \& Galletta 1994; Kuijken \&
Merrifield 1995; Merrifield \& Kuijken 1999; Bureau \& Freeman 1999).
Secular dynamical evolution has also been invoked to explain
correlations between disk and bulge properties (e.g., Courteau, de
Jong, \& Broeils 1996).

However, these evolutionary scenarios remain virtually untested for
active galaxies.  {\it Morphological} evidence for a barred
Seyfert galaxy with a boxy peanut-shaped bulge has recently been
presented by Quillen et al. (1997).  The present paper will discuss
the first unambiguous {\it kinematic} evidence for a bar potential in
two active galaxies with boxy bulges.  The objects we discuss --
the edge-on Sc galaxy NGC~3079 and Sb galaxy NGC~4388 --
show clear signs of nuclear activity at most
wavelengths. NGC~3079 is host to the most powerful windblown
superbubble known (Filippenko \& Sargent 1992; Veilleux et al. 1994,
hereafter VCBTFS). Infrared and radio measurements of this galaxy
suggest a nuclear starburst that coexists with an AGN
(e.g., Lawrence et al. 1985; Irwin \& Seaquist 1988; Haschick et
al. 1990; Irwin \& Sofue 1992; Baan \& Irwin 1995). The radio
morphology and optical/X-ray spectral properties of the nucleus
of NGC~4388 clearly point to a powerful AGN (e.g.,
Stone, Wilson, \& Ward 1988; Hummel \& Saikia 1991; Kukula et
al. 1995; Hanson et al. 1990; Iwasawa et al. 1997; Falcke et
al. 1998).  A complex of highly ionized line-emitting gas clouds
extends several kpc above the galactic plane
(e.g., Pogge 1988).  The origin of this extraplanar material has 
been discussed in detail by Veilleux et al. (1999; hereafter VBCTM).

The proximity of NGC~3079 (17.3 Mpc or 84 pc arcsec$^{-1}$ based on
Tully, Shaya, \& Pierce 1992) and NGC~4388 (16.7 Mpc or 81 pc
arcsec$^{-1}$ based on Yasuda, Fukugita, \& Okamura 1997) allows for
detailed structural studies.  The stellar bulge of
NGC~3079 presents a striking box/peanut shape at optical wavelengths
(Shaw, Wilkinson, \& Carter 1993 and references therein).
A stellar bar in NGC~3079 has been posited
(e.g., de Vaucouleurs et al. 1991), but such suggestions were often
based on the morphology of the central region rather than on
the kinematics. To date, the most detailed study of the stellar
kinematics in the bulge indicates cylindrical rotation that can be
explained without recourse to a non-axisymmetric distribution function
(Shaw et al. 1993). However, Merrifield \& Kuijken (1999) have
recently argued for a bar based on the peculiar [N~II]
emission-line profiles along the major axis of NGC 3079. The situation
for NGC~4388 is equally ambiguous. Recent K-band imaging by McLeod \&
Rieke (1995) revealed a boxy bulge, but the nearly edge-on aspect
of this galaxy prevented them from detecting the morphological
signature of a bar.

The two-dimensional velocity fields that we will present in this paper reveal
unambiguously a bar potential in both NGC~3079 and NGC~4388.  Contrary
to previous work based on long-slit spectra (e.g., Kuijken \& Merrifield
1995; Merrifield \& Kuijken 1999; Bureau \& Freeman 1999), our
detection of the bar potential in these two galaxies does not rely on
an inner Lindblad resonance ($x_2$-like orbits) or
strong bar-induced shocks that split emission-line
profiles when the bar is seen edge-on. 

Our paper is organized
as follows. In \S 2, we briefly describe the methods used to obtain
and reduce the near-infrared images and optical Fabry-Perot
observations.  Results on NGC~3079 and
NGC~4388 are discussed in \S 3 and \S 4, respectively. For each
galaxy, we first analyze the morphology of the stellar bulge and
its degree of boxiness. Next, we use the observed velocity
field of the ionized gas to argue for a bar potential.
Self-consistent dynamical interpretation of the disk velocity field
requires analysis of the surface photometry to constrain the mass
distribution, a task beyond the scope of this paper. The models
presented here are purely kinematical in nature, but they
show the clear kinematic signature of bar streaming
motions in both objects.  In \S 5, we discuss the implications of our
results using the predictions of bar evolutionary models, and
attempt to quantify how bars fuel the nuclear
activity. We summarize our results
in \S 6 along with future avenues of research.

\section{Observations and Data Reduction}

\subsection{Fabry-Perot Spectroscopy}

During the course of our spectroscopic survey of active galaxies with
extended line-emitting regions, we obtained a large Fabry-Perot
datacube (x, y, $\lambda$) of NGC~3079 that spans the H$\alpha$ +
[N~II] $\lambda\lambda$6548, 6583 emission-line complex, and two
datacubes of NGC~4388 centered on the H$\alpha$ and [O~III]
$\lambda$5007 emission lines.  Observational parameters of the
Fabry-Perot spectra are listed in Table 1. The observational setup and
reductions used to obtain and reduce these data have been detailed
elsewhere (VCBTFS, VBCTM, and Veilleux, Cecil, \& Bland-Hawthorn 1995;
hereafter VCB).  In both NGC~3079 and NGC~4388, the observed
emission-line profiles were parameterized by simple Gaussian
functions. Deviations from Gaussians will be discussed in \S 5.1.  The
best fits were determined by the least-$\chi^2$ method on spectrally
smoothed emission line profiles using a 1/4 -- 1/2 -- 1/4 spectral
filter (Hanning smoothing).  Spatial Gaussian smoothing with $\sigma$
= 1 pixel was used to improve the sensitivity to fainter features.
For the H$\alpha$ + [N~II] $\lambda\lambda$6548, 6583 complex in
NGC~3079, an iterative fitting method was used whereby all parameters
of the three Gaussian profiles were first left unconstrained (except
for the [N~II] $\lambda$6583/$\lambda$6548 ratio which was fixed to
its quantum value, 2.98; Osterbrock 1989). The continuum levels and
centroids of the H$\alpha$ and [N~II] profiles determined from this
first iteration were then used as input parameters for a second
iteration.

\subsection{Infrared Imaging}

Infrared images of NGC~3079 and NGC~4388 were obtained in the course
of imaging surveys of nearby galaxies by R. B. Tully and by
S. Courteau and J. Holtzman, respectively.  We thank our colleagues
for providing us with their data. On June 23 1993, the University of
Hawaii 2.2-meter telescope on Mauna Kea equipped with the NICMOS-3
camera (Hodapp, Rayner, \& Irwin 1992) was used to obtain a
K$^\prime$-band image of NGC~3079.  This imaging system uses a 256
$\times$ 256 pixel NICMOS-3 HgCdTe detector array and interchangeable
reimaging lenses to provide two spatial scales. The large angular size
of the program galaxies required the 2:1 reducing optics, resulting in
a field of $\sim$ 3$^\prime$ square at a scale of 0$\farcs$75
pixel$^{-1}$. The K$^\prime$ filter described in Wainscoat \& Cowie
(1992) minimized the thermal background. The integration time for each
frame was 1 minute, and the total exposure on NGC 3079 was 6 minutes.

The H-band image of NGC~4388 was obtained on April 28, 1994 using the
Cryogenic Optical Bench (COB) on the KPNO 2.1-meter telescope. This
system uses a 256 $\times$ 256 pixel InSb detector array, and provides
a field of $\sim$ 2$^\prime$ square at a scale of 0$\farcs$50
pixel$^{-1}$.  The total exposure on NGC~4388 was 1,000
seconds.  

Both image sets were reduced using standard techniques (e.g., Hodapp
et al. 1992; McCaughrean 1989), and were not flux calibrated because
we are only interested in the bulge/disk structure.

\section{Results on NGC~3079}

\subsection{Stellar Morphology}

The K$^\prime$-band image of NGC 3079 is shown in
Figure 1$a$, rotated to place the photometric 
major axis of the galaxy vertically.
This axis was found to be along PA$_{\rm maj}$ = 169$\arcdeg$ $\pm$
4$\arcdeg$, in good agreement with the results of Irwin \& Seaquist
(1991; PA$_{\rm maj}$ = 166$\arcdeg$).
Also shown are two simple photometric
models of the galaxy. In Figure 1$c$, the isophotes representing the
sum of an exponential disk and a spherically symmetric bulge are
shown, emphasizing the boxiness of the observed bulge.  An attempt is
made in Figure 1$b$ to model the observed isophotes more accurately
with a box-shaped structure of the form $I = I_{\rm box} + I_{\rm
disk}$ where
\begin{eqnarray}
I_{\rm box} = I_{b0}~(1 + \vert{R\over a}\vert^p + 
\vert{z\over b}\vert^p)^{-1/p}, \\
I_{\rm disk} = I_{d0}~{\rm exp}(-{R\over R_s})~{\rm 
exp}(-{\vert z\vert\over z_s}). 
\end{eqnarray}
This analytic formulation of the boxy structure was first introduced
by Pfenniger \& Friedli (1991). The parameters of the best fit model
are $I_{b0}/I_{d0}$ = 4.6, $a:b = 1:0.43$, $p$ = 3.5, $R_s$ = 3.0 kpc,
and $z_s$ = 380 pc (for an inclination of 82$\arcdeg$, \S 3.2.2).  For
comparison, the spherically symmetric bulge in Figure 1$c$ has
$I_{b0}/I_{d0}$ = 1.8, $b:a = 1:1$, and $p$ = 2.

The residual image after subtracting the boxy model
from the data is shown in Figure 1$d$. 
The striking X-structure in the central region
indicates a peanut-shaped bulge.  Although the boxiness of the
bulge in this galaxy has been known for some time (e.g., Shaw
1987; Young, Claussen, \& Scoville 1988; Shaw et al. 1993), our
infrared image emphasizes its importance by minimizing emission from the young
stellar disk and the effects of dust.
A detailed comparison of Figure 1$a$ and Figure 3 from VCB indicates
that the lower portions of the X-shaped line-emitting filaments
reported by VCB do {\it not} coincide spatially with the
peanut-shaped residual observed at infrared wavelengths.  The boxiness
of the infrared isophotes is therefore unlikely to be due to hot dust
or line emission from the X-shaped filaments.

Also visible in the residual image is a well-defined warp to
the north-west. A southern warp may also be present, but broader
spatial coverage is needed to see if it bends east
(``integral sign'' warp) or west (warp with mirror
symmetry with respect to the minor axis). The distributions of
line-emitting (VCB) and H~I gas (Irwin \& Seaquist 1991) strongly
favor the former orientation.

Finally, note the large residual nuclear core and disk
in Figure 1$d$. This nuclear disk coincides roughly
with the molecular disk and nuclear starburst in this galaxy
(Lawrence et al. 1985; Young, Claussen, \& Scoville 1988; Irwin \&
Sofue 1992; Sofue \& Irwin 1992; Baan \& Irwin 1995).

\subsection{Kinematics of the Gaseous Galactic Disk}

\subsubsection{General Description}

Figure 2$a$ reproduces
the distribution of line-emitting gas derived from our Gaussian
fits that was presented in VCB.
The velocity fields from these data are shown in Figures 2$b$ and
2$c$. The uncertainty on these velocities is
$\sim$ 35 km s$^{-1}$ in the brighter disk H~II regions,
but may be 2-3 times larger in the fainter material outside of
the disk. The H$\alpha$ and [N~II] $\lambda$6583 velocity fields shown
in Figures 2$b$ and 2$c$ generally agree within these errors.

In these figures a string of black dots traces the steepest gradient
through the observed velocity field (using the method described in
Bland 1986); this is the kinematic line of nodes.  Figure 3 shows the
H$\alpha$ rotation curve lay along this locus. It rises linearly in
the inner region and flattens to $\sim$ 245 $\pm$ 25 km s$^{-1}$ (or a
deprojected value of $\sim 250 \pm 25$ km s$^{-1}$ if $i$ =
82$\arcdeg$; \S 3.2.2) beyond a radius of $\sim$ 1 kpc.  A systemic
velocity of $\sim$ 1150 $\pm$ 25 km s$^{-1}$ is derived from this
rotation curve.  This value agrees well with estimates from other
optical datasets (e.g., 1177 km s$^{-1}$ from Humason et al. 1956;
1150 km s$^{-1}$ from Carozzi 1977) and CO spectra (e.g., 1150 km
s$^{-1}$ from Sofue \& Irwin 1992), but slightly exceeds the value
derived from HI data (e.g., 1120 km s$^{-1}$ from Rots 1980; 1125 km
s$^{-1}$ from Fisher
\& Tully 1981; 1118 $\pm$ 3 km s$^{-1}$ from Staveley-Smith \& Davies
1988; 1124 $\pm$ 10 km s$^{-1}$ from Irwin \& Seaquist 1991). This
apparent discrepancy is probably due to a slight asymmetry between the
inner optical/CO rotation curve ($R \la $ 8 kpc) and the
outer H~I rotation curve.  First noted by Sofue (1996), this asymmetry
may arise from a misaligned dark halo or tidal
interaction with nearby companions such as NGC~3073.
The systemic velocity derived from optical spectra of the stellar
bulge of NGC~3079 is closer to the HI value (1114 $\pm$ 9 km s$^{-1}$
from Shaw et al. 1993).

Well-defined `tongues' of highly redshifted (blueshifted) gas
extend immediately south (north) of the nucleus.  
Interestingly, these two `tongues' are slightly misaligned at the
nucleus.  The kinematic line of nodes also jumps
10$\arcdeg$ clockwise on either side of the nucleus
immediately outside the `tongues' ($R \ga$ 2.5 kpc;
Fig. 2), then twists slightly anti-clockwise.
This complex behavior may arise from a combination of
(1) a warp in the galactic disk, (2) patchy dust
obscuration, (3) streaming motion along spiral arms, and (4) eccentric
orbits aligned with a bar.  Although dust may contribute to the
slight large-scale differences between the present optical data and
the H~I data of Irwin \& Seaquist (1991),
the near-perfect bisymmetry of the velocity field near the center of
the galaxy strongly suggests that dust insignificantly affects the
observed velocity field. Similarly, the photometric warp detected on
the outskirts of the galaxy at optical, infrared, and radio
wavelengths, does not affect the kinematics of the gas inside a radius
of $\sim$ 8 kpc.  Finally, the coincidence between the clockwise shift
in the kinematic line of nodes and the anti-clockwise ``spiral arms''
seen on the K$^\prime$-band image of NGC 3079 (Fig. 1$a$) in both the
northern and southern sections of the disk is good evidence for
elliptic streaming through the spiral arms. However, this type of
streaming motion cannot explain the peculiarities of the velocity
field near the central region.

`Twisting' of the isovelocity contours in the central portion of
galaxies often signals a bar potential (e.g., Kalnajs
1978; Roberts, Huntley, \& van Albada 1979; Sanders \& Tubbs 1980;
Schwarz 1981).  The nuclear offset between the two `tongues' in
NGC~3079 can arise in this manner, as described next.

\subsubsection{Kinematic Models}

We began our analysis by exploring the parameter space for an
axisymmetric disk with inclination 50$\arcdeg$ $\le$ $i$ $\le$
90$\arcdeg$, kinematic major axis 160$\arcdeg$ $\le$ PA $\le$
200$\arcdeg$, and systemic velocity 1,000 $\le$ $V_{\rm sys}$ $\le$
1,300 km s$^{-1}$. This region of parameter space was selected based
on the results of previous optical and HI kinematic studies of
NGC~3079 (see references in the previous section).  A smoothed
(flux-weighted) version of the rotation curve derived along the
kinematic line of nodes (Fig. 3) was used for the analysis. When
searching for the best-fitting model, the whole disk out to R
$\approx$ 12 kpc was considered. Deviations from Gaussian profiles
were not considered in the following analysis (see \S 5.1).  Dust
obscuration, one likely source of profile asymmetry in this highly
inclined galaxy, is also not included in the models.

Figure 4 shows the best-fitting axisymmetric model ($i$ = 82$\arcdeg$,
PA = 169$\arcdeg$, and $V_{\rm sys}$ = 1,150 km s$^{-1}$). Figure 4$c$
shows the residuals after subtracting the model from our measured
H$\alpha$ velocity field.  We conclude that an axisymmetric model does
not fit the observed velocity field.

To look for the kinematic signature of a bar in NGC~3079, we
constructed more elaborate models of gas motions in the inner disk
that are similar to those described in Staveley-Smith et
al. (1990). We model the intrinsic gas orbits as nonintersecting,
inclined elliptical annuli that conserve angular momentum (so there
are no hydrodynamic shocks or gas flows near the corotation radius or
in the bar; Athanassoula 1992; Piner, Stone, \& Teuben 1995). The
ellipticity was held constant within a radius $R_b$, then was assumed
to decrease linearly out to a radius $R_d$ where the orbits are
circular.

The velocity field for the best fitting model is shown in Figure 4$b$;
model parameters are listed in Table 2. Our model invokes a bar with
moderately eccentric orbits ($e$ = $b/a$ = 0.7) with $R_b$ = 3.6 kpc
and $R_d$ = 6.0 kpc aligned along PA = 130$\arcdeg$ $\pm$ 10$\arcdeg$
intrinsic to the disk. This projects to PA = 97$\arcdeg$ on the sky
[using ${\rm tan}(PA_{\rm obs}) = {\rm tan}(PA_{\rm intrinsic}) {\rm
sec}~i$].  The position angle of the bar is well constrained by our
models. When the bar is close to the major or minor axis, one does not
get the twist of the line of node in the center, nor the NW-SE
bisymmetric structure at large radius. In this respect, the
intermediate angle bar models are a great success.  Compared to the
model with purely circular motions (Fig. 4$c$), the model that
incorporates elliptical streaming has significantly smaller residuals
in the central portion of the disk (Fig. 4$d$).  The dispersion in the
residuals within the inner disk (R $\la$ 6 kpc) is only 34 km s$^{-1}$
compared to 40 km s$^{-1}$ for models without the bar.

\section{Results on NGC~4388}

\subsection{Stellar Morphology}

The top panel of Figure 5 shows the H-band image of NGC 4388.  The
position of the photometric major axis of NGC 4388 derived from this
image is PA$_{\rm maj}$ = 90$\arcdeg$ $\pm$ 4$\arcdeg$. 
As for NGC~3079, we attempted to model the morphology
of NGC~4388 using either the sum of an exponential disk and a
spherically symmetric bulge (Fig. 5$c$) or the sum of an exponential
disk and a box-shaped bulge (eqns 1 \& 2; Fig. 5$b$). The second
solution clearly fits the data better.  The parameters
of this model are $I_{b0}/I_{d0}$ = 5.0, $a:b$ = 1:0.35, $p$ =
3.5, $R_s$ = 1.8 kpc, and $z_s$ = 0.32 pc (for inclination
--78$\arcdeg$, derived from our kinematic data; \S 4.2.2).

The residuals after subtracting this model from
the data are shown in Figure 5$d$.  Residual spiral arms
run east-west in the galactic plane and
coincide spatially with strings of HII regions in the H$\alpha$
emission-line image (see Fig. 1$a$ of VBCTM). The excess H-band emission
at these star-forming complexes is probably due to young
supergiant stars.  A bright fan-like
residual is also visible slightly south of the nucleus, coincident
with bright H$\alpha$ and [O~III] emission. Some of this
residual emission comes from the AGN in NGC~4388 (VBCTM).

\subsection{Kinematics of the Gaseous Galactic Disk}

\subsubsection{General Description}

The emission-line maps and velocity fields derived from the H$\alpha$
and [O~III] $\lambda$5007 data cubes were presented by VBCTM; the
velocity fields are reproduced in Figure 6.  Uncertainties range
from $\sim$20 km s$^{-1}$ in the bright line-emitting regions to $\ga$100
km s$^{-1}$ in the fainter areas. The ellipse superposed on
this figure differentiates between material in the disk and beyond.
A clear kinematic dichotomy is evident.  The models described
in the next section seek to reproduce the velocity field of the
disk material.  The kinematics of the extraplanar gas are discussed in
detail in VBCTM.

The velocity field of the [O~III]-emitting gas in the disk resembles
that of the H$\alpha$-emitting gas. It
is characterized by a large-scale east-west
gradient that indicates rotation in the galactic disk.  Figure 7
shows the rotation curve derived along the line of nodes of the
H$\alpha$ velocity field.  The rotation curve derived from our data is
consistent with earlier results (cf. Rubin, Kenney, \& Young 1997 and
references therein). The systemic velocity derived from this rotation
curve (2,525 $\pm$ 25 km s$^{-1}$) also agrees with published values
(2515 $\pm$ 7 km s$^{-1}$ from Helou et
al. 1981; 2,529 $\pm$ 3 km s$^{-1}$ from Corbin, Baldwin, \& Wilson
1988; 2,554 $\pm$ 39 km s$^{-1}$ from Ayani \& Iye 1989; 2,525 $\pm$
15 km s$^{-1}$ from Petitjean \& Durret 1993; 2,538 $\pm$ 26 km
s$^{-1}$ from de Vaucouleurs et al. 1991; 2,502 $\pm$ 10 km s$^{-1}$
from Rubin, Kenney, \& Young 1997).
`Twisting' of the isovelocity contours is clearly visible in the
central 1$\arcmin$ diameter of NGC~4388.  As for NGC~3079, the
near-perfect bisymmetry of the velocity field in the central region
of the galaxy strongly suggests that dust does not significantly
affect the observed velocity field there.  The kinematic models
described in the next section indicate that the velocity field is best
generated by a bar.

\subsubsection{Kinematic Models}

Our kinematic models seek to reproduce the H$\alpha$ velocity field.
In choosing the H$\alpha$ data for this analysis we have
tried to minimize effects associated with nuclear activity.
Dynamical processes such as entrainment by AGN-powered radio
jets generally have a stronger effect on the kinematics of the highly
ionized [O~III]-emitting gas than on those of the low-ionization
H$\alpha$-emitting material (e.g., Whittle et al. 1988).

The procedure used to find the best-fitting model for NGC~4388
followed the same steps as for NGC~3079. First, we explored the
parameter space for an axisymmetric disk with inclination
--90$\arcdeg$ $\le$ $i$ $\le$ --50$\arcdeg$ , kinematic major axis
along 70$\arcdeg$ $\le$ PA $\le$ 110$\arcdeg$, and a systemic velocity
of 2,400 $\le$ $V_{\rm sys}$ $\le$ 2,600 km s$^{-1}$ (this range in
the parameters brackets the results of previous kinematic studies; see
references in \S 4.2.1). Note that the negative inclination means that
the north rim of the disk is the near side.
This is consistent with the morphology
of the high-ionization gas as explained in VBCTM. Under these
conditions, the two spiral arms of bright HII regions are trailing
rotation, as generally arises in spiral galaxies.  To simplify
the analysis, we used a smooth (projected) rotation curve of the form:
\begin{eqnarray}
 V = V_0 {[1 - {\rm exp}(-\alpha~R)]\over [1 - {\rm exp}(-\alpha)]} + V_{\rm sys}, 
\end{eqnarray}
where $R$ is the galactocentric radius in arcseconds.
Deviations from Gaussians were not considered in our analysis (see \S 5.1).

Figure 8$a$ shows the best-fitting axisymmetric model ($i$ =
--78$\arcdeg$, PA = 90$\arcdeg$, $\alpha$ = 0.10 arcsec$^{-1}$, $V_0$
= 180 km s$^{-1}$ and $V_{\rm sys}$ = 2,525 km s$^{-1}$) and the
result of subtracting it from the H$\alpha$ velocity
field. The residuals in Figure 8$c$ are significant and
reveal substantial non-circular motion.
Variations in the rotation curve do not improve
significantly the quality of the fit.

Plausible dynamical origins for the non-circular motions in the disk
include forcing by a bar or oval distortion, density wave streaming
associated with the spiral arms, and disk warping. Our
spectra span only the inner 5 kpc of the disk, so the last effect
is unimportant.  A stellar bar has been suggested previously
based on the morphology of the central region, but
no consensus has emerged on its position angle
(Phillips \& Malin 1982: 30$\arcdeg$; Corbin, Baldwin, \&
Wilson 1988: 30$\arcdeg$; Colina et al. 1987: 130$\arcdeg$; see also
Rubin et al. 1997).  Figure 8$b$ presents the best fitting model that
incorporates elliptical streaming; Table 2 lists the model parameters.
The fit used a smooth rotation curve of form (3)
($V_0$ = 180 km s$^{-1}$, $\alpha$ = 0.10 arcsec$^{-1}$). Our
model involves a bar with radius $R_b$ = 1.5 kpc and highly eccentric
orbits ($e$ = 0.3) aligned along PA = 135$\arcdeg$ intrinsic to the
disk. This projects to PA = 100$\arcdeg$ on the sky.  The spiral-arm
effect arises because the gas orbits become increasingly less elliptical
from the ends of the bar to the edge of the disk ($e$ = 1 at the edge
of the disk, $R_d$ $\approx$ 5 kpc), and the intrinsic bar PA going
from 135$\arcdeg$ to 90$\arcdeg$. Elliptical streaming significantly
reduces the velocity residuals (Fig. 8$d$). The residuals present a
highly symmetric Gaussian distribution with a dispersion of only 20 km
s$^{-1}$. 
The fit is remarkably good considering that hydrodynamic effects
(e.g., strong shocks along the bar; Athanassoula 1992; Piner, Stone,
\& Teuben 1995) and dust obscuration are not modeled.

\section{Discussion}

\subsection{Additional Kinematic Evidence for Stellar Bars in
NGC~3079 and NGC~4388?}

Strong kinematic evidence for bar streaming motions in NGC~3079 and
NGC~4388 was presented in \S 3.2 and \S 4.2, respectively. The velocity
fields used for this analysis were derived by fitting Gaussians to the
emission-line profiles. Deviations from Gaussians
were detected in the inner disk of both galaxies. In NGC~3079, emission-line
profiles are split north of the nucleus (PA $\approx$ --10$\arcdeg$,
i.e. along the disk) out to a radius of $\sim$ 16$\arcsec$ (1.3 kpc),
but in a sector only $\sim$ 2 -- 3$\arcsec$ (170 -- 335
pc) wide. The degree of line splitting varies smoothly with radius, first
increasing monotonically to $\sim$275 km~s$^{-1}$ at 200 -- 300 pc
radii, beyond which it decreases. These anomalous line profiles were
described in Filippenko \& Sargent (1992; their Fig. 2) and VCBTFS
(their Fig. 11$a$).  Split line profiles were also detected by Irwin \& Sofue
(1992) in the inner H$_2$ molecular disk of this galaxy. Similar line
splitting with maximum amplitude $\sim$ 150 km s$^{-1}$ is observed
on both sides of the nucleus of NGC~4388 out to radii of $\sim$
10$\arcsec$ along the disk ($\sim$ 1 kpc; see also Iye \& Ulrich 1986;
Colina et al. 1987; Ayani \& Iye 1989; Veilleux 1991; Rubin et
al. 1997).

The origin of this line splitting is unclear. VCBTFS
tentatively interpreted the line splitting in the inner disk of
NGC~3079 as being due to the effects of the nuclear outflow on the
gaseous component of the galactic disk (see also Sofue \& Irwin 1992).
Rubin et al. (1997) have argued that the anomalous kinematics in the
inner disk of NGC~4388 indicates the presence of a discrete rapidly
rotating circumnuclear disk.  We argue that bar-induced
noncircular motions may also split the lines.
Kuijken \& Merrifield (1995) and, more recently,
Bureau \& Athanassoula (1999) have pointed out that the line-of-sight
velocity distribution of both the gaseous and stellar components in
barred potentials observed edge-on may be double-peaked and
characterized by a ``figure-of-eight'' variation with radius out to
roughly the end of the bar.  Their conclusions have since been
confirmed by hydrodynamical gas simulations (Athanassoula \&
Bureau 1999). Some of these simulations reproduce remarkably
well the velocity field of the inner disk of NGC~4388 (e.g., Fig. 3 of
Rubin et al. 1997).  The line splitting in this galaxy is symmetric
with respect to the nucleus and extends out to near the end of the
stellar bar.  The small extent of the line splitting in NGC~3079
relative to the size of the bar ($R_b$ = 3.6 kpc) and its asymmetry
with respect to the nucleus are more difficult to explain in this
scenario, but selective dust obscuration by material in the disk may
account for some of the asymmetry (see also long-slit observations of
this galaxy by Merrifield \& Kuijken 1999).

\subsection{The Bar -- Boxy Bulge Connection}

The co-existence of stellar bars and boxy bulges in NGC~3079 and
NGC~4388 brings additional observational support to evolutionary
models which posit that box/peanut-shaped bulges of disk galaxies
arise from a vertical instability in bars.  Given
bars in both galaxies, there is no need to invoke accretion
to explain the boxy bulges in these
galaxies (e.g., May, van~Albada, \& Norman 1985; Binney \& Petrou
1985; Rowley 1986, 1988). Note, however, that we cannot exclude the
possibility that each bar was created by a
galaxy interaction or merger and then evolved into a boxy bulge
through the bar-buckling instability (e.g., Noguchi 1987; Gerin,
Combes, \& Athanassoula 1990; Hernquist, Heyl, \& Spergel 1993; Mihos
et al. 1995; Miwa \& Noguchi 1998).

Evolutionary models which invoke resonant heating of the bar (Combes
et al. 1990; Pfenniger \& Norman 1990; Pfenniger \& Friedli 1991)
predict the formation of a boxy bulge over $\sim$10 bar
rotations.  The bulge may form even quicker if the bar is
subject to bending or fire-hose instabilities (Raha et al. 1991;
Merritt \& Sellwood 1994).  The bar rotation periods in NGC~3079 and
NGC~4388 can be derived assuming that the bars end near
the corotation radius and using the observed rotation
velocities there.  We find $\tau_{\rm
bar}$(NGC~3079) $\approx$ $\tau_{\rm bar}$(NGC~4388) $\simeq$ 1
$\times$ 10$^8$ yrs.
The boxy bulges in both NGC~3079 and NGC~4388 were therefore formed
over $\la$10$^9$ yrs.

The rate of bar heating/bulge formation is also a strong function
of the central mass concentration in these galaxies (e.g., Norman,
May, \& van Albada 1985; Hasan \& Norman 1990; Friedli \& Pfenniger
1991; Friedli \& Benz 1993, 1995; Hasan, Pfenniger, \& Norman 1993;
Norman, Sellwood, \& Hasan 1996; Merritt \& Quinlan 1998; Sellwood \&
Moore 1999). Large mass concentrations create inner-Lindblad
resonances (ILRs). ILRs form the anti-bar orbit families
$x_2$ and $x_3$, which thereby weaken the stellar bar.  

Simulations (Norman et al. 1996; and Merritt \& Quinlan 1998)
suggest that the bar may be destroyed within a fraction of its
rotation period if the core mass exceeds $\sim$ 1 -- 5\% of the
combined disk and bulge mass. Is this condition met in NGC~3079 and
NGC~4388?  An estimate of the core mass can
be derived assuming the nuclear activity is due to
accretion around a black hole of mass comparable to that of 
the galaxy core.  Recent compilations of black
hole masses (Kormendy et al. 1997; Richstone 1998; Magorrian
et al. 1998) suggest a rough proportionality between black hole
and bulge masses with $M_{\rm BH} \simeq 0.005 M_{\rm
bulge}$. The constant of proportionality may depend slightly
on morphological type, perhaps increasing from late- to early-type
galaxies (e.g., $M_{\rm BH} \sim 2.5\% M_{\rm bulge}$ in the S0
galaxies NGC~3115 and NGC~4342; Merritt 1998), but this variation is
of little importance here because we are dealing with two late-type
galaxies.  In this case, we get
\begin{eqnarray}
\eta \equiv {M_{\rm core} \over M_{\rm bulge} + M_{\rm disk}}
\simeq {M_{\rm BH} \over M_{\rm bulge} + M_{\rm disk}} 
\simeq 0.005 {M_{\rm bulge} \over M_{\rm bulge} + M_{\rm disk}}
\la 0.005,
\end{eqnarray}
implying that the central black holes in these galaxies do not play a
significant role in the destruction of the bars.  The very presence of
AGN in NGC~3079 and NGC~4388 may {\em
require} $\eta$ to be smaller than the critical value for rapid bar
dissolution because destruction of the bar would end mass accretion
onto the nucleus (see Merritt 1998).

The short time scales that we derive suggest that we are unlikely to
have caught the bars in the act of forming boxy bulges.  This is
probably also the case for most barred spirals with boxy bulges, given
the large fraction of disk galaxies with strong bars ($\sim$ 35\%;
e.g., Shaw 1987; Sellwood \& Wilkinson 1993; Mulchaey \& Regan 1997)
and box-peanut bulges (20 -- 45\%; e.g., Jarvis 1986; Shaw 1987; de
Souza \& dos Anjos 1987; Dettmar \& Barteldrees 1988, 1990; see Dwek
et al. 1995, Kuijken 1996, and references therein for a discussion of
our own Galaxy).
Numerical simulations suggest that bars can indeed persist after the
epoch of boxy bulge formation (e.g., Miller \& Smith 1979; Pfenniger
\& Friedli 1991; but see Raha et al. 1990 for a counterexample).  It
is also possible that bar formation, dissolution, and bulge building
recurs in disk galaxies. Open-box simulations of disk galaxies with
nearby companions show that tidal encounters can repeatedly form a bar
throughout the life of a disk galaxy (e.g., Sellwood \& Moore 1999).

\subsection{Stellar Bars and Nuclear Activity}

Stellar bars in NGC~3079 and NGC~4388 do not
necessarily assure efficient fueling of the nuclear starbursts/AGN.
Our kinematic models neglect hydrodynamical effects and therefore
cannot constrain the bar-induced mass inflow rates in these
galaxies. Several factors are important in determining if these
bar-induced mass inflow rates suffice to power the observed
nuclear activity.  Possibly relevant parameters include the strength
and length of the bar, the gas mass-fraction and morphological type of
the host galaxy, the star formation efficiency, the age of the bar,
and the existence of ILRs near the nucleus.  In
this section, we review briefly each of these possibilities in the
context of NGC~3079 and NGC~4388.

\subsubsection{Length and Strength of the Bar}

Surveys of barred galaxies find that those currently displaying the
highest star formation activity have both strong and long bars (e.g.,
Martin 1995; Martinet \& Friedli 1997). Here, we follow Martinet \&
Friedli and define strong and long bars as having deprojected bar axis
ratios $(b/a)_i \le 0.6$ and relative lengths $2 L_i/D_{25} \ge 0.18$
where $L_i$ is the deprojected bar length and $D_{25}$ is the galaxy
diameter at 25 mag arcsec$^{-2}$. The high inclinations of NGC~3079
and NGC~4388 prevent us from determining photometrically the lengths
and axis ratios of these bars, but the results from our analysis of
the velocity fields (Table 2) constrain these values. Assuming $L_i
\approx R_b$ and using $D_{25}$ = 40 kpc for NGC~3079 and 27 kpc for
NGC~4388 (de Vaucouleurs et al. 1991), we get relative lengths of
$\sim$ 0.18 and $\sim$ 0.11 for NGC~3079 and NGC~4388,
respectively. The bar in NGC~3079 therefore is a borderline case while
that of NGC~4388 appears to be a short bar.  The strengths of the bars
may be estimated using the derived eccentricity (= $b/a$) of the gas
orbits in the inner ($R \le R_b$) portion of the bars. The bar in
NGC~4388 would therefore be considered a strong bar while that in
NGC~3079 is weak.  Note, however, that using the eccentricity of the
gas orbits rather than that of the stellar orbits probably
overestimates the strength of the stellar bar by factors of 2 -- 3
(e.g., Friedli \& Benz 1993), so that even the bar in NGC~4388 may be
relatively weak. The modest size and weakness of the stellar bars
in both galaxies may prevent efficient fueling of their active nuclei.

\subsubsection{Gas Mass Fraction, Morphological Type, and 
Star Formation Efficiency}

That not all galaxies with strong and long bars are forming stars furiously
(Martinet \& Friedli 1997) indicates that
other parameters also affect the star formation rate in barred
galaxies.  The numerical simulations of Friedli \& Benz (1995)
suggest that the overall rate of star formation increases with
gas mass-fraction of the galaxy.  These authors found that a
larger gas mass-fraction pushes the threshold for star formation to
larger radii in the disk, thereby increasing the overall star
formation rate.  However, feedback between star formation
and the release of mechanical energy from supernovae keeps the rate of star
formation in the central regions relatively constant.  This may
explain why the correlation between gas mass-fraction and {\em
nuclear} star-formation is more evident in early-type barred spirals,
where star formation rates and feedback from supernovae are more
modest (e.g., Hawarden et al. 1986; Devereux 1987; Dressel 1988;
Arsenault 1989; Huang et al. 1996).

For NGC~3079, little can therefore be said on the importance of the gas 
mass-fraction in determining bar-induced fueling.
However, it is clear that the nuclear starburst in this galaxy has the 
potential to be long-lived regardless of the mass inflow
rate induced by the bar. Indeed, given the current star formation rate
in the nucleus ($\sim$ 10~$\beta^{-1}~{\rm M}_\odot~{\rm
yr}^{-1}$ where $\beta < 1$ is the fraction of the bolometric
luminosity of NGC~3079 radiated by stars; VCBTFS) and the total mass
of atomic and molecular gas observed in the galaxy {\it core} (Irwin
\& Seaquist 1991), star formation will deplete gas in the nuclear region
of NGC~3079 in $>$ $\beta$ 10$^9$ yrs. Bar-induced
inflow would increase this time.

The situation in NGC~4388 is less clear because the nuclear
activity appears to be powered by an AGN rather than a starburst. We
are unaware of numerical simulations that address
the effects of gas mass-fraction or morphological type on
the level of AGN activity. 

\subsubsection{Age of the Bar}

Self-consistent evolutionary models of barred galaxies by Friedli \&
Benz (1995), Martinet \& Friedli (1997) and Martin \& Friedli (1997)
suggest that both the total and nuclear star formation rates in
galaxies with strong (weak) bars peak $\la$ 1 (2) Gyr after
the bar instability starts.  The {\em relative} importance of
nuclear star formation increases more than tenfold
over that time span. According to strong (weak) bar
simulations, young barred galaxies [$\la$ 0.5 (1.0) Gyrs; Type A in
the nomenclature of Martin \& Friedli 1997] are characterized by
chains of bright HII regions near strong shocks along the bar with
no star formation activity in the nucleus, while old barred galaxies
[$\ga$ 0.8 (1.6) Gyrs; Type C] shows the opposite distribution.
The discussion in \S 5.3.1 suggests that the time scales for
the weak bar simulations are more likely to apply to the bars in
NGC~3079 and NGC~4388.  In principle, the star formation distribution
in these bars can therefore constrain the age of the bars
and determine if the bar is old enough to fuel the observed
nuclear activity. Unfortunately, both NGC~3079 and
NGC~4388 are highly inclined spirals and patchy dust obscuration
affects the brightness of HII regions in these galaxies.
Qualitatively, we find that the bar region in NGC~3079 presents a
larger number of bright HII regions than that of NGC~4388, therefore
suggesting that the stellar bar in NGC~3079 is less evolved than the
one in NGC~4388.

Perhaps it is more instructive to invert this argument and
attempt to constrain the age of the bar by assuming that the nuclear
activity in NGC~3079 and NGC~4388 is induced by the bar. Then,
the age of the bar should equal the sum of the starburst or AGN
lifetime and the delay between bar formation and the onset of
nuclear activity. This situation may be particularly relevant to
NGC~3079, where a nuclear starburst seems to power the
kpc-scale windblown superbubble (VCBTFS). The dynamical time scale of
the superbubble ($\sim$ 10$^6$ yrs; VCBTFS) and starburst age
(10$^7$ -- 10$^8$ yrs) determined from the optical spectrum of the
nucleus (VCBTFS) are both considerably shorter than the predicted
delay between bar formation and the onset of the starburst ($\sim$
10$^9$ yrs). If the starburst in NGC~3079 was triggered
by the bar, this would imply that the bar is $\sim$ 10$^9$
yrs old.  The age of the bar equals or exceeds
the timescale that we derived in \S 5.2 for the formation
of the boxy bulge in NGC~3079.

The situation is more complex in NGC~4388, where the nuclear activity
is almost certainly due to mass accretion onto a supermassive black
hole rather than from a starburst. The ionized gas detected above the
disk of NGC~4388 (VBCTM) may be used to constrain the age of the AGN
in this galaxy. This extraplanar material appears to be outflowing
from the nucleus with a characteristic dynamical time scale of $\sim$
2 $\times$ 10$^7$ yrs. If no other outflows have occured
in NGC~4388, the dynamical time scale of the extraplanar
material constrains the age of the AGN to a few $\times$ 10$^7$
yrs. Once again, this is much shorter than the predicted
delay between the epoch of bar formation and the onset of
starburst activity predicted by numerical simulations.  However, more
relevant to NGC~4388 is the delay between the epoch of bar
formation and the onset of AGN activity. This delay depends critically
on poorly constrained factors including star formation efficiency,
the conversion efficiency of the kinetic energy injected from supernovae
and stellar
winds into gas motion (if both these efficiencies are large, less
material will be available to fuel the AGN), and the time needed to
transport material from the kpc-scale starburst down to the sub-pc
scale of the AGN accretion disk. Simulations of
AGN fueling by bars (e.g., Heller \& Shlosman 1994) suggest that intense
starburst activity always precedes or coincides with AGN activity.
Our data can therefore only place a lower limit of $\sim$
10$^9$ yrs on the age of the bar in NGC~4388. This is again
consistent with the time scale that we derived for the formation of
its boxy bulge.

\subsubsection{Inner Lindblad Resonances}

In the self-consistent evolutionary models discussed in the previous
section, the ILRs appear at the end of the simulations once
sufficient mass has inflowed. ILRs have
important consequences on the fueling of the nuclear starburst and
perhaps the AGN.  ILRs correspond to regions of orbit
crowding associated with higher gas concentrations and sometimes
accompanied by shocks (e.g., Athanassoula 1992; Piner et
al. 1995). ILRs in barred galaxies often form
a gas ring nearby (e.g., Telesco, Dressel,
\& Wolstencroft 1993). Gas flows down the dust
lane of the bar to the ILR, then most sprays back into the bar region
at the point of contact of the nuclear ring and the dust lane (e.g.,
Binney et al. 1991; Piner et al. 1995; Regan et al. 1997).

A galaxy can have zero, one, or two ILRs depending on its
rotation curve and therefore on the detailed mass distribution.  For a
weak bar rotating with angular velocity $\Omega_b$ in a galaxy in
which stars at a galactocentric radius $R$ (cylindrical
coordinates) orbit with angular velocity $\Omega(R) = V(R)/R$, an ILR
at that radius for which the bar potential perturbs, or
drives, the stars at a frequency equal to their natural (or epicyclic)
frequency $\kappa(R)$; i.e.  $\Omega_b = \Omega(R) - \kappa(R)/2$ at
ILR where (e.g., Binney \& Tremaine 1987):
\begin{eqnarray}
\kappa(R)^2 = {2V \over R} ({V \over R} + {dV \over dR}).
\end{eqnarray}
Our discussion in \S 5.3.1 suggests that this weak bar approximation
is realistic for both galaxies.  Unfortunately, we have no easy way to
estimate $\Omega_b$ in NGC~3079 and NGC~4388.  We assume that the
corotation radius, $R_{\rm CR}$ where $\Omega_b = \Omega(R_{\rm CR})$,
coincides with the ends of the bar (i.e., $R_{\rm CR} \simeq R_b$),
and derive $\Omega_b$ $\simeq$ 60 km s$^{-1}$ kpc$^{-1}$ in NGC~3079
and $\Omega_b$ $\simeq$ 105 km s$^{-1}$ kpc$^{-1}$ in NGC~4388. Next,
we smoothed the rotation curves to calculate $\kappa(R)$ from equation
(5).  For NGC~3079, we follow the procedure of Bland-Hawthorn,
Freeman, \& Quinn (1997) and fit the observed rotation using a
dynamical model which comprises a Freeman exponential disk and a dark
halo of the form
\begin{equation}
\rho_h = \rho_\circ (1 + r^2/r_a^2)^{-1}
\end{equation}
with the rotation curve $V(r)$ given by
\begin{equation}
\label{arctan}
V^2 = V_\infty^2 [1 - ({{r_a}\over{r}})\ \tan^{-1} ({{r}\over{r_a}})].
\end{equation}
This fitting procedure was not used for NGC~4388 because
the nuclear outflow affects the
velocity field north-east of the nucleus. Instead we deprojected equation
(3) using $V_0$ = 180 km s$^{-1}$ and $\alpha$ = 0.1 arcsec$^{-1}$.

The results of our calculations are shown in the lower panels of
Figures 9 and 10. The adopted rotation curves are shown in the top
panels of these figures.  We find that neither galaxy has an
ILR.  This result is consistent with the presence of large amounts of
gas in the cores ($\la$ 1$\arcsec$) of these objects.  The absence of
an ILR may help the bar-induced mass inflows continue
down to nuclear scales to fuel the AGN and/or nuclear starburst. Note,
however, that the converse may be generally false: results from
numerical simulations suggest that ILRs do not necessarily prevent
transport of material toward the nucleus (e.g., Friedli \& Benz
1993).  Consequently, the absence of an ILR in galactic nuclei is
probably not a necessary condition for efficient fueling of AGN and
nuclear starbursts.

\section{Summary}

In this paper, we have presented the first direct kinematic evidence
for bar potentials in two active galaxies with box/peanut-shaped
bulges. The complete two-dimensional coverage of our Fabry-Perot spectra
has allowed us to detect the kinematic signature of the bar potential
without an ILR or strong bar shocks.
Our results provide observational support for
bar evolutionary models where boxy bulges represent a critical
transitional phase in the evolution of stellar bars into spheroidal
bulges. We compared the predictions of these models with our data to
determine that the boxy bulges in NGC~3079 and NGC~4388 probably
formed over a time scale $\la$ 10$^9$ yrs. This short
time scale, if typical of all barred galaxies, is difficult to
reconcile with the high frequency of bars and boxy bulges in galaxies
unless (1) bars survive the epoch of boxy bulge formation,
or (2) bar and boxy bulge formation recur in the life of
disk galaxies.  Perhaps minor mergers reform a bar
and feed new fuel to the host galaxy.

Stellar bars in NGC~3079 and NGC~4388 provide a mechanism for fueling
the nuclear activity in these galaxies. However, we find no {\em
direct} evidence that mass inflows induced by the bars suffice to
power the nuclear activity in these objects. The bars in NGC~3079 and
NGC~4388 are rather short and weak and may not be dynamically
important enough to trigger the required mass inflows. Using the
velocity fields and bar kinematics derived from the Fabry-Perot data,
we find that both galaxies lack inner Lindblad resonances. This may
explain the large quantities of gas found in the nuclei of these
galaxies, and the fueling of these active galaxies down to a scale of
10 pc. However, our evaluation of the bar-induced fueling of the
nuclear activity in these galaxies is severely limited by the current
lack of simulations that attempt to relate the mass inflow rates
outside $\sim$ 10 pc with the processes affecting gas dynamics closer
in. Such simulations would be particularly relevant to NGC~4388 where
the nuclear activity is probably due to an AGN (evidence for both a
nuclear starburst and an AGN exists in NGC~3079).  A proper treatment
of the MHD effects associated with the AGN will be needed to address
this important issue.

From an observational standpoint, more efforts should be made to
search for small ($\la$ 10 pc) nested bars in nearby active galaxies
and to map the velocity field of the gaseous and stellar components in
this region to constrain the dynamical importance of bars in fueling
nuclear activity. As recent examples of such work, we note the
intriguing results on Seyfert galaxies obtained by Regan \& Mulchaey
(1999) and Martini \& Pogge (1999).  Using the WFPC2 and NICMOS
instruments on HST, they found a paucity of nuclear bars in the galaxies
studied but several ``nuclear mini-spirals''.  Follow-up kinematical
data may establish if such structures can fuel the AGN.

\clearpage

\acknowledgments

We thank S. Courteau, J.  Holtzman, J. Huang, and R. B. Tully for
acquiring and reducing the near-infrared images of NGC~3079 and
NGC~4388 used in the present paper.  We also thank the referee,
R. Pogge, for constructive comments that helped improve the paper. SV
is grateful for partial support of this research by a Cottrell
Scholarship awarded by the Research Corporation, NASA/LTSA grant NAG
56547, NSF/CAREER grant AST-9874973, and Hubble fellowship
HF-1039.01-92A awarded by the Space Telescope Science Institute which
is operated by the AURA, Inc. for NASA under contract No. NAS5--26555.
JBH acknowledges partial support from the Fullam award of the Dudley
Observatory.

\clearpage

\begin{table*}
\caption{Fabry-Perot Observations}
\begin{center}
\begin{tabular}{llll}
\tableline
\tableline
\noalign{\vskip 7.5 pt}
                     & NGC~3079           & NGC~4388           & NGC~4388\\
\tableline
\noalign{\vskip 7.5 pt}
Emission Lines       & H$\alpha$ + [N~II] $\lambda\lambda$6548, 6583 &
H$\alpha$ & [O~III] $\lambda$5007\\
Dates of Observations&  1990 March 14 -- 17  & 1990 May 31 \& June 1 & 1992 March 12\\
Total Exposure Time  &  20 hrs            & 6.3 hrs            & 7.0 hrs\\
Telescope            &  CFHT 3.6m         & UH 2.2m            & UH 2.2m\\
Field of View        &  4$\farcm$6        & 10$\arcmin$        & 10$\arcmin$\\
Pixel Scale          &  0$\farcs$57       & 0$\farcs$66        & 0$\farcs$85\\
Spectral Resolution  &  70 km s$^{-1}$    & 40 km s$^{-1}$     & 60 km
s$^{-1}$\\
Number of Spectra    &  150,000           & 5,000              & 7,000\\
\noalign{\vskip 7.5 pt}
\tableline
\end{tabular}
\end{center}
\end{table*}

\clearpage

\begin{table*}
\caption{Morphological and Kinematical Parameters of NGC~3079 and NGC~4388}
\begin{center}
\begin{tabular}{llll}
\tableline
\tableline
\noalign{\vskip 7.5 pt}
Component &Parameter & NGC~3079           & NGC~4388\\
\tableline
\noalign{\vskip 7.5 pt}
Disk &$V_{\rm sys}$  &1,150 $\pm$ 25 km s$^{-1}$ & 2,525 $\pm$ 25 km s$^{-1}$\\
     & $V_{\rm max}$ (deprojected) & 250 $\pm$ 25 km s$^{-1}$&180 $\pm$ 25 km s$^{-1}$  \\
     &P.A. (major axis)    &169$\arcdeg$ $\pm$ 4$\arcdeg$& 90$\arcdeg$ $\pm$ 4$\arcdeg$\\
     &Inclination $i$ &82$\arcdeg$ $\pm$ 4$\arcdeg$& --78$\arcdeg$ $\pm$ 4$\arcdeg$\\
     &Scale length $R_s$ & 3.0 kpc & 1.8 kpc\\
     &Scale height $z_s$ & 0.38 kpc & 0.32 kpc\\
     &Adopted rotation curve&Smoothed version & Equation (3) with \\
     &                      &of observed rotation& $V_0$ = 180 km s$^{-1}$ and \\
     &                      &curve               &$\alpha$ = 0.1 arcsec$^{-1}$\\
Bulge&Axis ratio $a:b$& 1:0.43& 1:0.35\\
     &Boxiness $p$ & 3.5& 3.5\\
     &Intensity relative to disk $I_{b0}/I_{d0}$ & 4.6 & 5.0\\
Bar  &P.A. (intrinsic to disk) &130$\arcdeg$ $\pm$ 10$\arcdeg$& 135$\arcdeg$ $\pm$ 15$\arcdeg$\\
     &Inner radius $R_b$ &3.6 kpc& 1.5 kpc\\
     &Orbit eccentricity $R \le R_b$& 0.7& 0.3\\
     &Angular velocity $\Omega_b = \Omega(R_b)$& 60 km s$^{-1}$ kpc$^{-1}$& 105 km s$^{-1}$ kpc$^{-1}$\\
     &Outer radius $R_d$ &6.0 kpc& 5.1 kpc\\
\noalign{\vskip 7.5 pt}
\tableline
\end{tabular}
\end{center}
\end{table*}

\clearpage

\begin{figure}
\caption{ Stellar morphology of NGC 3079. ($a$) K$^\prime$ band image of
NGC 3079, ($b$) model consisting of an exponential disk and a box-shaped
bulge, ($c$) model consisting of an exponential disk and a spherical
bulge, ($d$) residuals resulting from the subtraction of ($b$) from ($a$).
}
\end{figure}

\begin{figure}
\caption{The left panel shows the distribution of the H$\alpha$
emission in NGC~3079.  The middle and right panels show the velocity
fields derived from H$\alpha$ and [N~II] $\lambda$6583, respectively.
Each panel shows the central 125$\arcsec$ $\times$ 280$\arcsec$ (10.5
kpc $\times$ 23.5 kpc) of NGC~3079 with north at the top and east to the
left. The line of nodes derived from each of these velocity fields is shown
superposed as a series of black dots.}
\end{figure}

\begin{figure}
\caption{ The top panel shows the rotation curve of NGC 3079 derived
along the position of the line of nodes of the H$\alpha$ velocity
field.  The velocity range at each radius represents $\pm$ 1
$\sigma$. The full line is the smoothed, flux-weighted rotation curve
used in the two-dimensional fits to the velocity field (Fig. 4). The
dashed line corresponds to the systemic velocity, 1,150 km s$^{-1}$.
The velocities along the minor axis are shown in the lower panel.}
\end{figure}

\begin{figure}
\caption{ Predicted velocity fields for ($a$) an axisymmetric disk, ($b$)
a disk with bar streaming motions.  The residual maps (observed
H$\alpha$ velocity field of NGC~3079 $-$ model) for the axisymmetric
and bar models are presented in panels ($c$) and ($d$), respectively.
The parameters of the bar model are listed in Table 2. The bar model
fits the central disk considerably better.  }
\end{figure}

\begin{figure}
\caption{ Stellar morphology of NGC 4388. (a) H-band image of NGC
4388, (b) model consisting of an exponential disk and a box-shaped
bulge, (c) model consisting of an exponential disk and a spherical
bulge, (d) residuals resulting from the subtraction of (b) from (a).
}
\end{figure}

\begin{figure}
\caption{ Velocity field of the line-emitting gas in NGC 4388.  (top)
[O~II] $\lambda$5007; (bottom) H$\alpha$. North is at the top and east
to the left. }
\end{figure}

\begin{figure}
\caption{ Rotation curve of NGC 4388. The data points were derived along
the position of the line of nodes of the H$\alpha$ velocity field. The
rotation curve was stopped where outflowing motions north-east of the
nucleus become significant.  The dotted line shows the systemic
velocity of the galaxy, 2,525 km s$^{-1}$.  The solid lines represent
two attempts to fit these data using equation (3). The thin solid line
uses $V_0 = 120$ km s$^{-1}$ and $\alpha$ = 0.18 arcsec$^{-1}$ while
the thick line uses $V_0$ = 180 km s$^{-1}$ and $\alpha$ = 0.10
arcsec$^{-1}$. The two-dimensional bar model subtraction shown in
Figure 8$d$ clearly demonstrates that the rotation curve represented by
the thick line is a better representation of the data.  The deviations
from the modeled rotation curve are due to streaming motions.  }
\end{figure}

\begin{figure}
\caption{ Predicted velocity fields for ($a$) an axisymmetric disk, ($b$)
a disk with bar streaming motions.  The residual maps (observed
H$\alpha$ velocity field of NGC~4388 $-$ model) for the axisymmetric
and barred disk models are presented in panels ($c$) and ($d$),
respectively.  The parameters of the barred disk model are listed in
Table 2. The barred disk model fits the observations considerably
better.  }
\end{figure}

\begin{figure}
\caption{ The top panel shows the deprojected rotation curve used
in our fit of the Fabry-Perot data of NGC~3079 with a bar model
(solid line in the top panel of Figure 3) and the best-fit rotation
curve calculated using a Freeman exponential disk (dashed curve) and
an isothermal dark halo (dotted line). See text for detail. The lower
panel shows the behavior of $\Omega$ (thick solid line), $\Omega \pm
\kappa/2$ (dotted lines) and $\Omega_b$ (dashed line) derived
using the best-fit rotation curve in the upper panel.  No ILR appears
to be present in NGC 3079.}
\end{figure}

\begin{figure}
\caption{ Same as Figure 9 but for NGC 4388. The best-fit rotation
curve from Figure 7 (the thick solid line in that figure) was used for
the calculations after deprojection.  No ILR appears to be present in
NGC 4388.}
\end{figure}

\clearpage

%\setcounter{figure}{0}
%\begin{figure}
%\plotfiddle{Fig1.eps}{6.00in}{0}{90}{90}{-290}{-150}
%\caption{}
%\end{figure}

%\clearpage

%\begin{figure}
%\plotthree{Fig2a.eps}{Fig2b.eps}{Fig2c.eps}
%\caption{}
%\end{figure}

%\clearpage

\setcounter{figure}{2}
\begin{figure}
\plotone{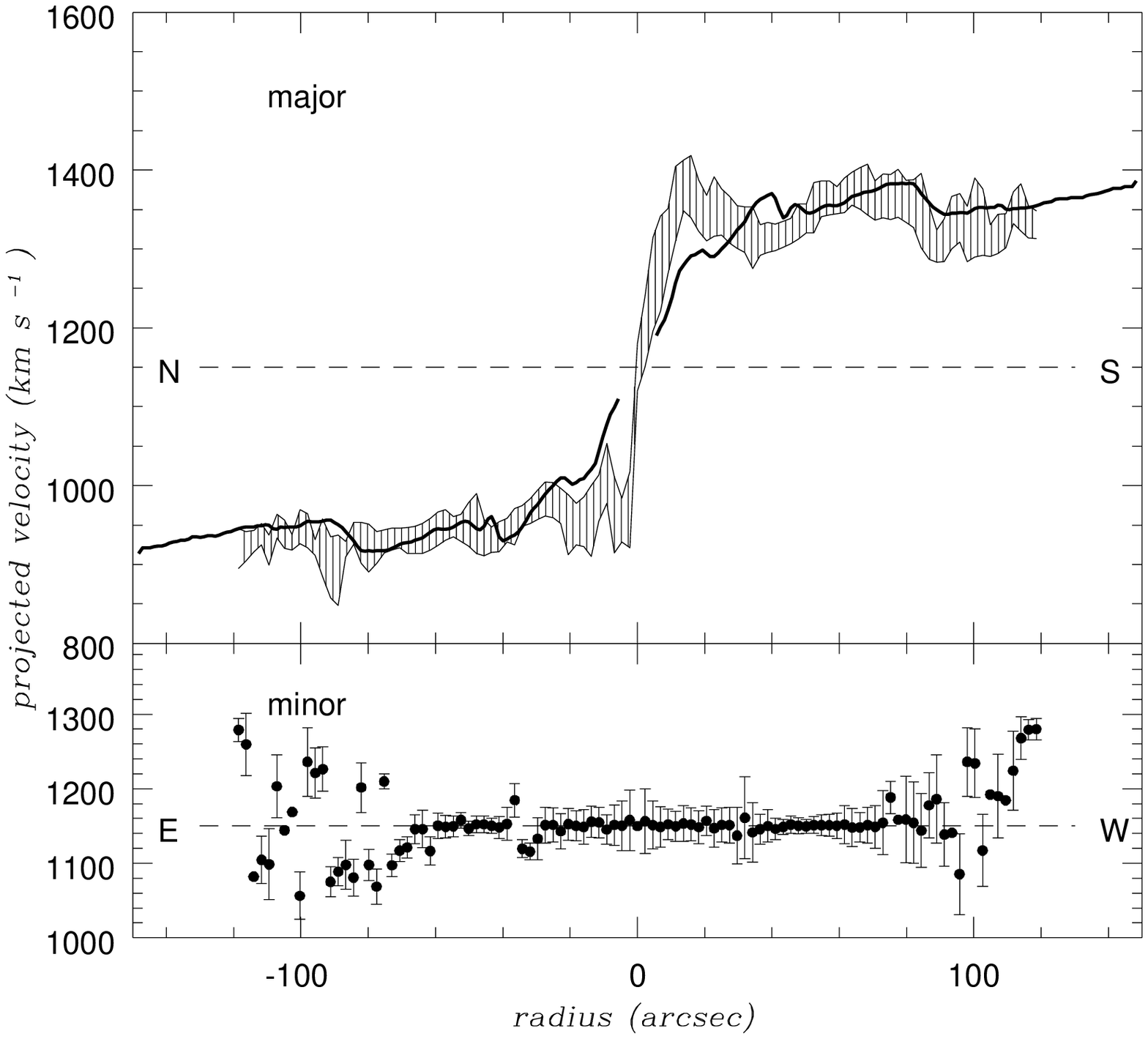}
\caption{}
\end{figure}

\clearpage

%\begin{figure}
%\plottwo{Fig4a.eps}{Fig4b.eps}
%\caption{(a) and (b)}
%\end{figure}

%\setcounter{figure}{3}
%\begin{figure}
%\plottwo{Fig4c.eps}{Fig4d.eps}
%\caption{(c) and (d)}
%\end{figure}

%\clearpage

%\begin{figure}
%\plotfiddle{Fig5.eps}{6.00in}{0}{120}{120}{-180}{-300}
%\caption{}
%\end{figure}

%\clearpage

%\begin{figure}
%\plotone{Fig6.eps}
%\caption{}
%\end{figure}

%\clearpage

\setcounter{figure}{6}
\begin{figure}
\plotone{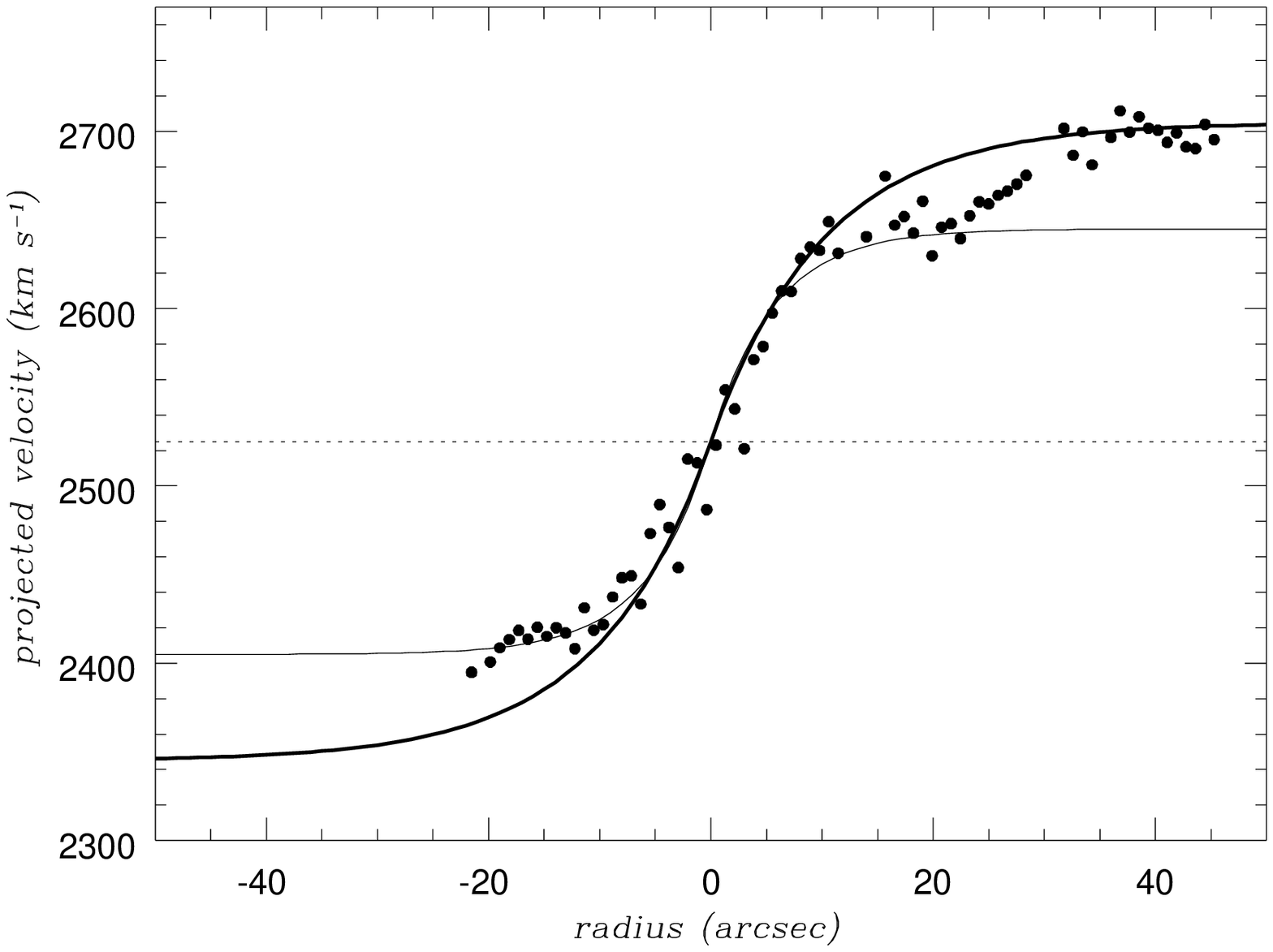}
\caption{}
\end{figure}

\clearpage

%\begin{figure}
%\plottwo{Fig8a.eps}{Fig8b.eps}
%\plottwo{Fig8c.eps}{Fig8d.eps}
%\caption{}
%\end{figure}

%\clearpage

\setcounter{figure}{8}
\begin{figure}
\plotone{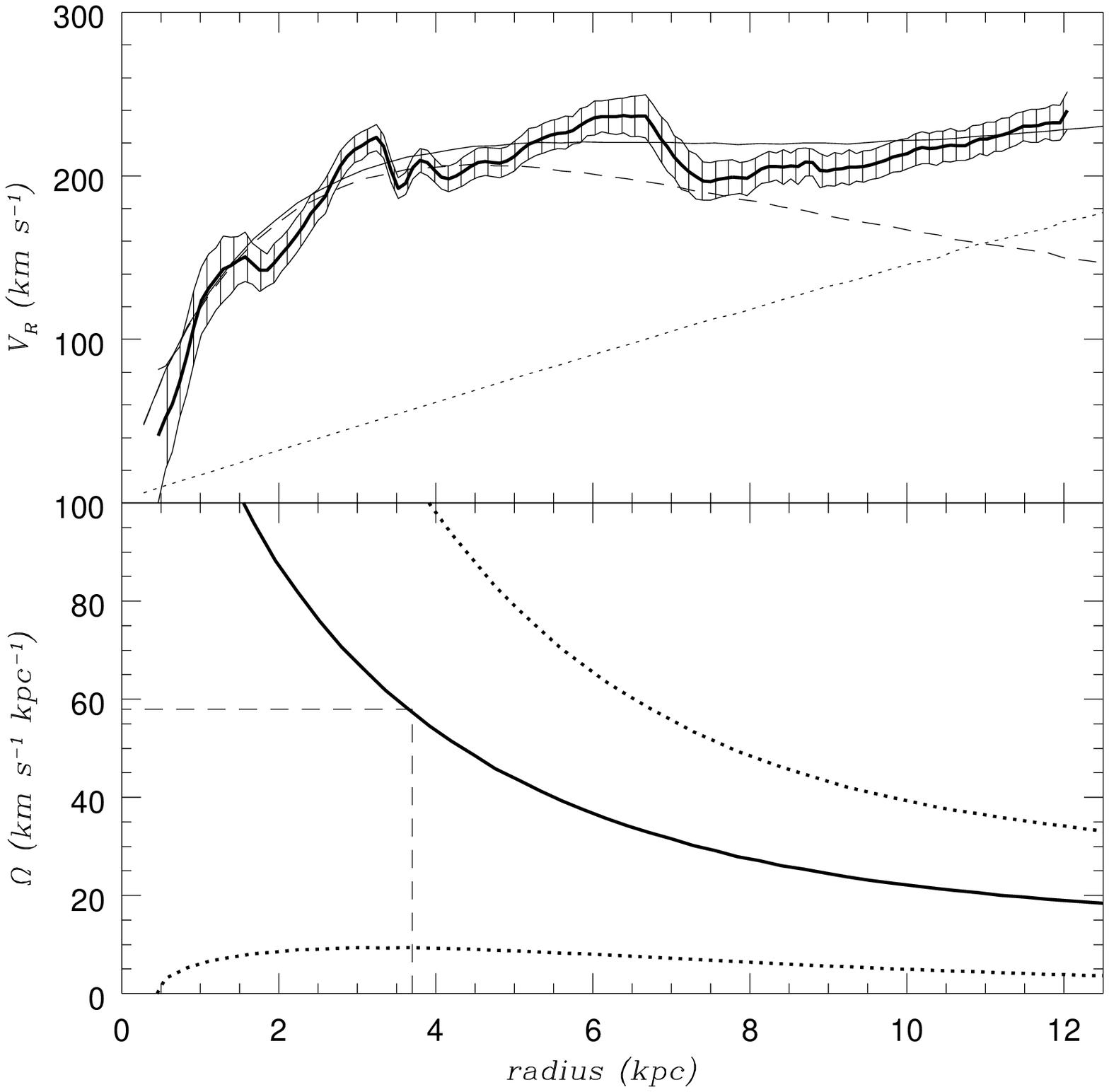}
\caption{}
\end{figure}

\clearpage

\begin{figure}
\plotone{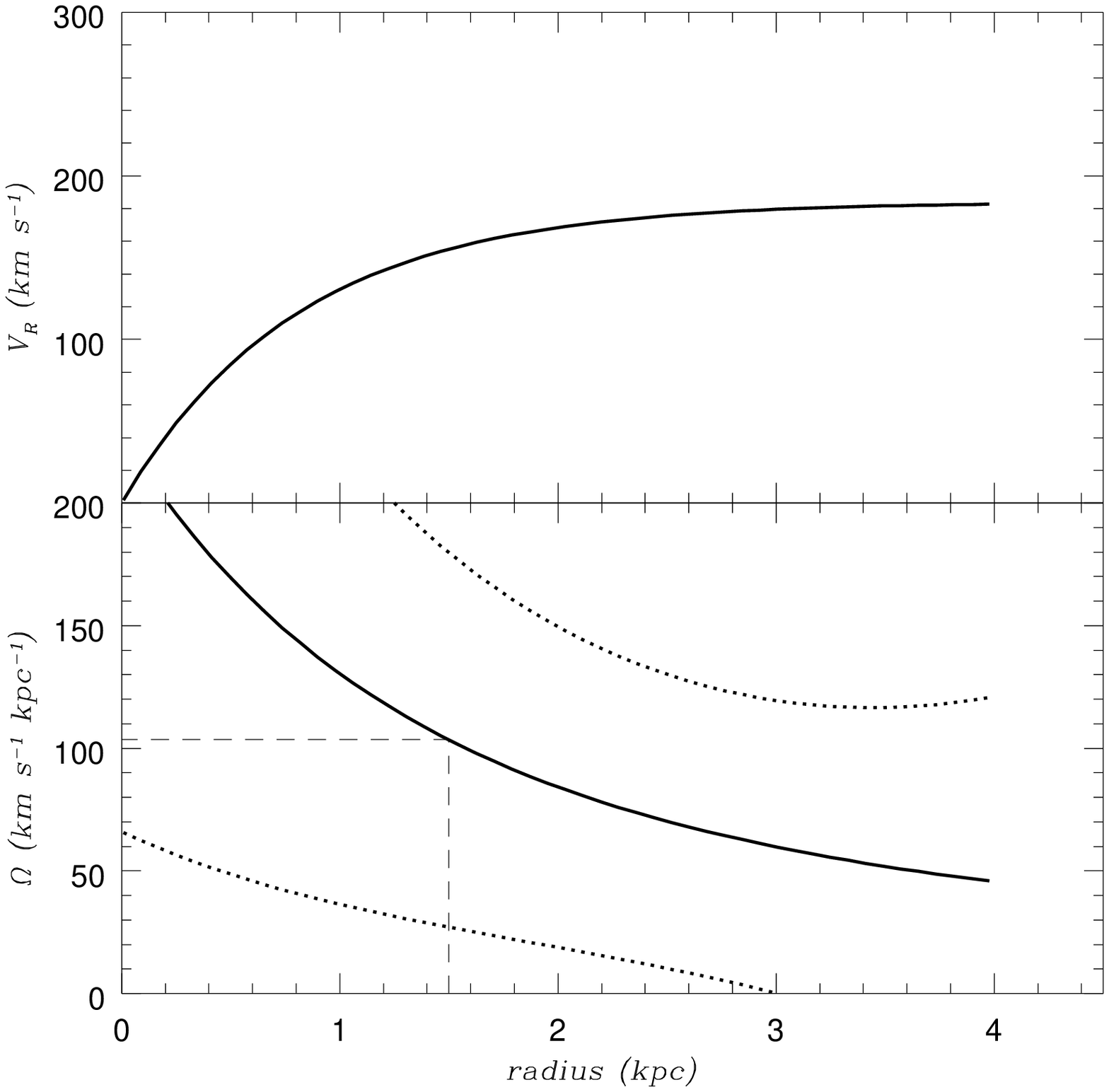}
\caption{}
\end{figure}

\end{document}